\documentclass{article}
\usepackage{graphicx}
\def\p{\partial}
\def\s{\sigma}

\def\G{\Gamma}
\def\g{\gamma}

\def\d{\delta}
\def\de{\delta}
\def\D{\Delta}
\def\ld{\lambda}
\def\Ld{\Lambda}
\def\L{\Lambda}
\def\ep{\epsilon}
\def\e{\eta}

\def\om{\omega}

\def\b{\beta}

\def\a{\alpha}

\def\pdellx'{\frac{\partial}{\partial x'}}
\def\pdellw'{\frac{\partial}{\partial w'}}

\newcommand{\be}{\begin{equation}}
\newcommand{\ee}{\end{equation}}
\def\bed{\begin{displaymath}}
\def\eed{\end{displaymath}}
\def\bea{\begin{eqnarray}}
\def\eea{\end{eqncrray}}
\def\[{$$}
\def\]{$$}

\begin{document}
\title{A model of unified quantum \\ chromodynamics and  Yang-Mills gravity  }

\author{HSU Jong-Ping \\
Department of Physics,
University of Massachusetts Dartmouth \\
North Dartmouth, MA 02747-2300, USA}


\maketitle
{\small   Based on a generalized Yang-Mills framework, gravitational and strong interactions can be unified in analogy with the unification in the electroweak  theory.   By gauging $T(4) \times [SU(3)]_{color} $ in flat space-time, we have a unified model of chromo-gravity with a new tensor gauge field, which couples universally to all gluons, quarks and anti-quarks.  The space-time  translational gauge symmetry assures that all wave equations of quarks and gluons reduce to a Hamilton-Jacobi equation with the same `effective Riemann metric tensors'  in the geometric-optics (or classical) limit.    The emergence of effective metric tensors in the classical limit is essential for the unified model to agree with experiments.  The unified model suggests that all gravitational, strong and electroweak interactions appear to be  dictated by gauge symmetries in the generalized Yang-Mills framework.} 
\bigskip
\bigskip
 
\bigskip

\bigskip

\noindent
\bigskip

A theory of  `Yang-Mills gravity' has been formulated recently on the basis of external translational gauge symmetry within the framework of flat space-time with arbitrary coordinates.\cite{1,2}  The basic assumption is that the gravitational action, involving the translational gauge covariant derivative $\p_{\mu} + g\phi_{\mu}^{\nu}T_{\nu}$,  is invariant under the local space-time translations.  Such a translational gauge symmetry for gravity dictates the presence of a symmetric tensor field associated with $T(4)$ generators $T_{\mu}$.  The generators $T_{\mu}$ of the external space-time translational group $T(4)$ have the representations $T_{\mu}=\p /\p x^{\mu}$.  In contrast, the generators of internal groups, such as $SU(2) \times U(1)$ and $SU(3)$ for electroweak theory and quantum chromodynamics respectively,\cite{3,4,5} have constant matrix representations. The gauge transformations of internal gauge groups correspond to phase changes.  However, `external' $T(4)$ gauge transformations correspond to scale changes.  We may remark that beside the usual phase change, a scale change also occurs in a gauge theory based on the de Sitter group.  In this case, in order to restore gauge symmetry of a Lagrangian involving fermions, it is necessary to introduce `scale gauge fields,' in addition to the usual phase gauge fields (i.e., the Yang-Mills fields).\cite{6} 
The usual Yang-Mills framework with internal groups (and the associated vector gauge potential fields) can be generalized to include external space-time groups (and the associated tensor gauge potential fields).   This generalization enables us to unify gravitational interaction and strong interaction in analogy to that in the Weinberg-Salam theory.  Such a general framework may be called the  generalized Yang-Mills framework. 

 The flat space-time  translational gauge symmetry in the unified model assures that all wave equations of quarks and gluons reduce to a Hamilton-Jacobi equation, which explicitly shows the emergence of the same `effective Riemann metric tensor'  in the geometric-optics (or classical) limit.  This gives a new and interesting picture of the physical world.  Namely, in the presence of $T(4)$ gauge fields, the classical objects and light rays exhibit  motions as if they were in a `curved space-time.'  However, the real physical space-time for fields, quantum particles and their interactions is flat, i.e. the Riemann-Christoffel curvature tensor vanishes.  The $T(4)$ gauge symmetry in flat space-time is the key for the unified model to be consistent with experiments and for us to quantize the gravitational field.
Possible  relations between space-time symmetry and gravity were recognized and explored in the literature.\cite{7,8,9}  But their formulations of gravity were based on curved space-time or different dynamical fields, and their gravitational actions differ from that in the present unified model. 
 Moreover, the space-time translational symmetry and its relation to gravity have been discussed by many authors.\cite{9,10,11,12,13} For example, there is a formulation called teleparallel gravity,\cite{13} which is based on the translational gauge symmetry on a flat space-time with a torsion tensor.  In teleparallel gravity, a flat connection with torsion makes translations local gauge symmetries and its curvature tensor is the torsion field.   However, there are basic differences between Yang-Mills gravity and teleparallel gravity.   For example,  they have different gauge potentials, different gravitational actions and different physical implications.
  
To show the significance of the space-time translational gauge symmetry, let us  gauge the symmetry groups $T(4) \times [SU(3)]_{color}$ in the generalized  Yang-Mills framework.  We obtain a unified `quantum chromogravity' (QCG).  Similar to $U(1)$ symmetry in quantum electrodynamics, the unified model involves $T(4)$ symmetry, so that the $T(4)$ group generators must appear in the Lagrangian through a gauge covariant derivative.  Suppose the operators  $T_{\nu}$ denote the generators (in general) of the flat space-time translation group and satisfy $[T_{\mu}, T_{\nu}]=0 $.  We shall use the representation $T_{\mu}=\p_{\mu}$ for our discussions.   In the generalized Yang-Mills framework, the $T(4)$ gauge covariant derivative is assumed to be obtained by the following replacement,\cite{1,2} 
\be
\p_{\mu} \to \p_{\mu} + g\phi_{\mu}^{\nu}T_{\nu} = J_{\mu}^{\nu}\p_{\nu},  \ \ \ \  \ \ \ \
J_{\mu}^{\nu}= (\d_{\mu}^{\nu} + g\phi_{\mu}^{\nu}),
\ee
in inertial frames with the Minkowski metric tensor $\e_{\mu\nu}=(1,-1,-1,-1)$.     We see in (1) that a mixed tensor field $\phi_{\mu}^{\nu}$ is naturally associated with the $T(4)$ generator $T_{\mu}=\p_{\mu}$, which has the dimension of 1/length, so that the associated coupling constant $g$ has the dimension of length.  Furthermore, the $T(4)$ coupling terms in the unified model will contain the factor $g\phi_{\mu}^{\nu}T_{\nu}$, which does not involve $i=\sqrt{-1}$, so that there is only one single force between matter-matter, matter-antimatter, and antimatter-antimatter, in contrast to the electromagnetic  force with a dimensionless coupling constant.\cite{1,2}  These properties are just right for gravity.  For a general frame of reference (inertial and non-inertial)  with the Poincar\'e metric tensors $P_{\mu\nu}$, the form of the gauge covariant derivative (1) still holds, provided we replace  $\p_{\mu}$ by the covariant derivative $D_{\mu}$ associated with the metric tensor, $P_{\mu\nu}$.  In the limit of zero acceleration, the metric tensor $P_{\mu\nu}$ reduces to the Minkowski metric tensor $\eta_{\mu\nu}$.\cite{1}   Therefore, the unified gauge covariant derivatives ${\bf d}_{\mu}$ in `quantum chromogravity' for general frames are assumed to be  
\be
{\bf d}_{\mu} \equiv D_{\mu} + g\phi_{\mu}^{\nu}D_{\nu} +ig_{s}{G_{\mu a}}\frac{{ \ld_a}}{2},  \ \ \ \
\ee
where $\ld_a$ ($a$=1,2,3...8,) are the fundamental representation of the eight generators of color $SU(3)$, and $g_s$ is the usual dimensionless strong coupling constant in quantum chromodynamics.   The fields $G_{\mu}^a$ are the color gauge vector potentials, i.e., the gluon fields.\cite{3}   

In order to see the gauge curvatures  associated with the unified gauge covariant derivative ${\bf d}_{\mu}$, let us calculate the commutators $[{\bf d}_{\mu} ,{\bf d}_{\nu}]$.  We obtain
\be
[{\bf d}_{\mu} ,{\bf d}_{\nu}]=  C_{\mu\nu\s} D^{\s}  + ig_s {\bf G}_{\mu\nu a} \frac{\ld_a }{2}. 
\ee
The new $[SU(3)]_{color}$ gauge curvature (i.e., the color gauge field strength tensor) in the presence of the gravitational gauge potential is given by  ${\bf G}_{a\mu\nu}$,
\be
{\bf G}_{a \mu\nu} =  J_{\mu}^{\s}D_{\s}G_{a \nu} - J_{\nu}^{\s}D_{\s}G_{a \mu} - g_{s}f_{abc}G_{b \mu}G_{c \nu}, 
\ee
where $a,b,c=1,2,...8, $ and the structure constants $f_{abc}$ are completely anti-symmetric. The $T(4)$ gauge curvature $C_{\mu\nu\a}$ is given by
\be
C_{\mu\nu\a}= J_{\mu\s}D^{\s} J_{\nu\a}-J_{\nu\s} D^{\s} J_{\mu\a}=\D_{\mu}J_{\nu\a}-\D_{\nu}J_{\mu\a}=-C_{\nu\mu\a},
 \ee 
 $$\D_{\mu}=J_{\mu\nu}D^{\nu}, \ \ \ \ \ \   J_{\mu\nu}= P_{\mu\nu} + g\phi_{\mu\nu}, \ \ \ \ \ \   \phi_{\mu\nu} = \phi_{\nu\mu}=P_{\nu\ld}\phi_{\mu}^{\ld}.$$
Equations (2)-(5) hold in arbitrary coordinates within flat space-time.  This $T(4)$ gauge curvature $C_{\mu\nu\a}$ differs completely from the Riemann-Christoffel curvature of space-time.  The burning question is whether the $T(4)$ gauge curvature $C_{\mu\nu\a}$ in flat space-time can be consistent with gravitational experiments.   The answer turns out to be affirmative, as we shall see below.

 
Within the generalized Yang-Mills framework with the gauge curvatures in (4) and (5), the action functional and the Lagrangian of the unified quantum chromogravity (QCG) model in general frames are assumed to take the quadratic form in gauge theory, 
\be
S_{QCG}=\int L_{QCG} d^4 x,  \ \ \ \ \ \ \  L_{QCG}=L_{G}+L_{\phi q}
\ee
 \be
L_{G}= -\frac{1}{4}{\bf G}^{a\mu\nu}{\bf G}^{a}_{\mu\nu}\sqrt{-P}, \ \ \ \ \ \   L_{\phi q}=[L_{\phi}+ \overline{q}(i\G^{\mu}{\bf d}_{\mu}- M)q]\sqrt{-P},     
\ee 
$$L_{\phi}= \frac{1}{4g^2}\left (C_{\mu\nu\a}C^{\mu\nu\a}- 2C_{\mu\a}^{ \ \ \  \a}C^{\mu\b}_{ \ \ \  \b} \right), \ \ \    \{\G_\mu, \G_\nu\} = 2 P_{\mu\nu},    $$
$$
  \G_\mu=\g_a e^a_\mu,
    \ \ \     
\{\g_a, \g_b\} = 2 \eta_{ab},  \ \ \   \eta_{ab} e_\mu^a e_\nu^b =
P_{\mu\nu}, \ \ \  D_{\mu} q=\p_{\mu} q,
$$
where $P = det P_{\mu\nu}$ and  $q's$ denote collectively spinor quarks with three colors and six flavors.\cite{3}   We observe that there are two independent quadratic forms that one can construct with the $T(4)$ gauge curvature $C_{\mu\nu\a}$.\cite{1}   It is gratifying that their linear combination, as shown in $L_{\phi}$, turns out to be consistent with gravitational experiments.     
   
 The $[SU(3)]_{color}$ gauge transformations in the unified model are just those in the usual quantum chromodynamics, provided the gravitational gauge fields $\phi_{\mu\nu}(x)$ are assumed to transform trivially under $[SU(3)]_{color}$ because they are not affected by the strong force.  In contrast, the  $T(4)$ gauge transformations are different because the local space-time translations are given by 
\be
x^{\mu} \to x'^{\mu} = x^{\mu}+\Ld^{\mu}(x),
\ee
where $\Ld^{\mu}(x)$ are arbitrarily infinitesimal vector functions.  However, this local translations (8) with an arbitrary $\Ld^{\mu}(x)$ are also  general infinitesimal coordinate transformations.  Thus, the $T(4)$ gauge transformations for vector or tensor fields (in a general frame of reference with arbitrary coordinates) are formally the same as the Lie variation of tensors.\cite{1} 
 For a space-time-dependent tensor
$Q^{\mu_1 ...\mu_m}_{\a_1 ...\a_n} (x)$ in the
 action (6), including $P_{\mu\nu}$, the $T(4)$ gauge transformations are assumed to be
$$
 Q^{\mu_1 ...\mu_m}_{\a_1
...\a_n} (x)^{\$} 
=\frac{\p x'^{\mu_1}}{\p
x^{\nu_1}}... \frac{\p
x'^{\mu_m}}{\p x^{\nu_m}}
\frac{\p x^{\b_1}}{\p x'^{\a_1}} ... \frac{\p x^{\b_n}}{\p
x'^{\a_n}} 
$$
\be
\times  \left(1 - \L^\lambda (x)
\p_\ld \right)Q^{\nu_1 ...\nu_m}_{\b_1 ...\b_n} (x),
\ee
where $\mu_1, \nu_1, \a_1,\b_1,$ etc. are flat space-time indices.  Here, we have suppressed indices of internal groups because they do not change the external gauge transformations (9). 
For example, the $T(4)$ gauge transformations of a scalar function $S(x)$ and a covariant tensor $T_{\mu\nu}$ are as follows:
\be
 S(x)^{\$}=S(x) - \L^\ld(x) \p_\ld S(x),  \ \ \ \   S(x)=q(x), \overline{q}(x), \psi(x), \ or  \ \Phi(x),
 \ee 
$$  T_{\mu\nu}(x)^{\$}=T_{\mu\nu}(x) - \L^\ld(x) \p_\ld T_{\mu\nu}(x) -
T_{\mu\a}(x) \p_\nu
\L^\a(x) - T_{\a\nu}(x) \p_\mu \L^\a(x),   $$
where $ T_{\mu\nu}= J_{\mu\nu}$ or
$P_{\mu\nu}$.    
 As usual, (Lorentz) spinor field $\psi$, (Lorentz) scalar field 
$\Phi$ and quark fields ($q(x)$ and $\overline{q}(x)$) are treated as `coordinate scalars.'  
 The covariant derivative of a covariant vector $V_{\nu}$, i.e.,  $D_{\mu}V_{\nu}$, transforms like a covariant tensor $T_{\mu\nu}$.   We have seen from (10) that
the external $T(4)$ gauge transformations of a spinor field $\psi$ and scalar field correspond to scale changes rather than phase changes.   

In the unified model, the translational symmetry in flat space-time is intimately related to the conservation of the energy-momentum tensor.  Otherwise, the quantum chromogravity will not be completely satisfactory if it is not compatible with the established conservation law.  The generalized Yang-Mills framework based on flat space-time with arbitrary coordinates is just general enough to realize this symmetry-conservation-law connection in all physical frames of reference.\cite{14,1}  It appears that if one goes beyond this generalized Yang-Mills framework and works on a more general framework such as the curved space-time, this symmetry-conservation-law connection may be lost in general.  We stress that the conservation laws are the physical essence of the symmetry properties, as revealed through Noether's theorem.

Based on the translational   gauge transformation (9), we can show 
that  the scalar Lagrangian $L_{Q}=L_{QCG}/ \sqrt{-P}$ and $\sqrt{-P}$ transform as follows, 
$$
L_{Q} \to (L_{Q})^{\$} =L_{Q} - \L^\ld (\p_\ld L_{Q}). 
$$
 \be
\sqrt{-P} \to \sqrt{-P^{\$}}
= \left[(1-\L^\s \p_\s)\sqrt{-P} \ \right](1 - \p_\ld
\L^\ld), \ \ \ \   P = det P_{\mu\nu}.
\ee
Thus,  we obtain
\be
\int L_{QCG} d^4 x  \to
\int \left[\sqrt{-P}L_{Q} -  \p_\ld \left(\L^\ld L_{Q} \sqrt{-P}  \ \right)\right]d^{4}x = \int 
L_{QCG} d^4 x 
\ee
The divergence term in (12) does not contribute to the action and, hence, does not affect the field equations.  
Thus, we have shown that, although the QCG Larangian is not gauge invariant, the action $S_{QCG}$ given by (6) is invariant under the $T(4)$ gauge transformations.   

In the unified model, the translational gauge symmetry in flat space-time implies an effective curved space-time for the motion of all classical objects.  The $T(4)$ gauge symmetry dictates that all quarks, antiquarks and gluons are univesally coupled to gravity through the $T(4)$ gauge covariant derivatives $\D_{\mu}=J_{\mu\nu}\p^{\nu}$ in the Lagrangian.    We examine the classical limit (i.e., the geometric-optics limit) of quark and gluon wave equations.
 First, let us consider only a quark and gravitational field in inertial frames (i.e., $P_{\mu\nu}= \e_{\mu\nu}$), for simplicity.  The Lagrangian for a quark $q$ in the presence of the gravitational gauge potential takes the form
\be
L_{q}=
\frac{i}{2}[\overline{q} \g_\mu \Delta^\mu q - (\Delta^\mu
\overline{q}) \g_\mu  q] -
m\overline{q} q,
\ee
$$\g_{\mu}\g_{\nu}+\g_{\nu}\g_{\mu}=2\e_{\mu\nu}, \ \ \ \ \ \    \D_{\mu}=J_{\mu\nu}\p^{\nu},  \ \ \ \ \  J_{\mu\nu} = J_{\nu\mu},$$
where we have symmetrized the quark Lagrangian, so that quarks and antiquarks have the same coupling to gravity. The wave equation for a quark can be obtained, at least formally,
\be
i\g_\mu \Delta^\mu q - m q + \frac{i}{2} \g_{\mu}[\p 
_{\nu}(J^{\mu\nu} )] q = 0, 
\ee
although a free quark and gluon do not exist because of the permanent quark confinement.
Using the limiting expression  for the fermion field
$q= q_{o}exp(iS)$, where $q_{o}$ is a constant spinor, and the properties that $g\phi_{\mu\nu}$ and $g\p_\a \phi_{\mu\nu}$ are extremely small for gravity, we can derive the quark equation
\be
[\g_{\mu} J^{\mu\s} \p_{\s} S + m]q_o  = 0, \ \ \ \ \ \   J^{\mu\s}= (\e^{\mu\s} + g\phi^{\mu\s}).   
\ee
In the classical limit, the momentum $\p_{\mu} S$ and mass $m$ are 
large quantities.\cite{1,15}
To eliminate the spin variables, we multiply a 
factor $(\g_{\s} J^{\s\mu} \p_{\mu} S - m)$ to  (15).  
With the help of the anti-commutation relation for $\g_{\s}$ in (13), equation (15) leads to the Hamilton-Jacobi 
equation with an `effective Riemann metric tensor' $G_{\mu\nu}$,
\be
G^{\mu\nu}(\p_{\mu}S)(\p_{\nu}S) - m^{2} = 0,  \ \ \ \ \ \ \    G^{\mu\nu} = \e_{\a\b} J^{\a\mu} J^{\b\nu}. 
\ee
This is the equation of the motion of a classical particle in the presence of the 
gravitational tensor field $\phi^{\mu\nu}$.  

For the classical limit of gluon wave equation in the presence of  gravitational field, let us consider the wave equation of a gluon field coupled only to the gravitational tensor field $\phi_{\mu\nu}$.  
In inertial frames, the gluon wave equation can be derived from the Lagrangian (7), 
\be
\p_{\mu}(J^{\mu}_{\s}{\bf G}_a^{\s\nu}) - g_s f_{abc}G_{b \mu}{\bf G}_c^{\mu\nu} =0.
\ee
Using the limiting expression $G_a^{\mu}=\epsilon_a^{\mu} exp(iS')$ and choosing a `Lorentz gauge' for the gluon field $\p_{\mu}G_a^{\mu} = 0$, the wave equation (17) for the massless gluon reduces to
\be
G^{\mu\nu}(\p_{\mu}S')(\p_{\nu}S')  = 0,  \ \ \ \ \ \    G^{\mu\nu} = \e_{\a\b} J^{\a\mu} J^{\b\nu},  
\ee
in the geometric-optics limit, where the wave vector  $\p_{\mu}S'$ is large.  This equation (18) for massless gluon field is the same as the eikonal equation for the light ray.\cite{1}    

   The gravitational implications for observable phenomena described quantum chromodynamics in microscopic world are negligible.  So, let us concentrate on experimental implications of  gravity for macroscopic world in the unified model.   We consider the action $S_{\phi q}$ involving only $\phi_{\mu\nu}$ and quark fields $q(x)$ in general frames with Poincar\'e metric tensor $P_{\mu\nu}$:
\be
S_{\phi q} = \int (L_{\phi q}  + L_{gf})d^4 x,  \ee
\be
L_{gf}=\frac{1}{2g^2}\e_{\a\b}\left[\p_\mu J^{\mu\a} - \frac{1}{2}\e^{\a\ld}\p_{\ld} J^\mu_\mu\right]
\left[\p_\nu J^{\nu\b} - \frac{1}{2}\e^{\b\ld}\p_{\ld} J^\nu_\nu\right],
\ee
where $L_{\phi q}$ is given in (7) with the $T(4)$ gauge curvature $C_{\mu\nu\a}$ given by (5).  We have included a   gauge-fixing term $L_{gf}$ specified by (20) involving ordinary partial derivative to break the $T(4)$ gauge symmetry so that the solution of gauge field equation is well-defined.  The reason for including $L_{gf}$ is  that field equations with gauge symmetry are known to be not well defined in general and that it is a nuisance to find explicit solutions of such field equations without having a gauge-fixing term.\cite{16}  

The quark fields play the source for producing a gravitational potential field $\phi_{\mu\nu}$.  The $T(4)$ gravitational field equation for symmetric tensor field, $\phi_{\mu\nu}=\phi_{\nu\mu}$ can be derived from (19),
\be
H^{\mu\nu} +  A^{\mu\nu}=    \frac{g^2}{2} Sym  \ \left[ \overline{q} i\G^\mu D^\nu q -
i(D^\nu \overline{q}) \G^\mu q \right]\equiv g^2 T^{\mu\nu} , 
\ee
$$
H^{\mu\nu} = Sym 
\left[D_\ld (J^{\ld}_\rho C^{\rho\mu\nu} - J^\ld_\a 
C^{\a\b}_{ \ \ \ \b}P^{\mu\nu} + C^{\mu\b}_{ \ \ \ \b} J^{\nu\ld})  \right.
$$
\be
\left. - C^{\mu\a\b}D^\nu J_{\a\b} + C^{\mu\b}_{ \ \ \ \b} D^\nu J^\a_\a -
 C^{\ld \b}_{ \ \ \ \b}D^\nu J^\mu _\ld\right],
\ee
\be
A^{\mu\nu} =Sym \left[ \p^\mu \left(\p^\ld J_\ld{^\nu}  - \frac{1}{2} \p^\nu 
J \right)  - \frac{\e^{\mu\nu}}{2}  \left(\p^\a \p^\ld J_{\ld\a}  
 - \frac{1}{2} \p^\a \p_\a J \right)\right], 
\ee
where $D_{\mu}q=\p_{\mu}q$ and $J= J^\ld_\ld$.  The symbol `Sym' in Eqs. (21)-(23) denotes that $\mu$ and $\nu$ should be made symmetric.  Note that (23) always involves ordinary derivatives (even when one uses spherical coordinates) because it is derived from the gauge-fixing terms defined by the Lagrangian $L_{gf}$ in (20).   

The gauge-fixing terms used in a previous discussion\cite{1} involved the covariant derivative $D_{\mu}$.  They are the same when one works in inertial frame with $\e_{\mu\nu}$ for quantization of fields.  Nevertheless, the previous gauge-fixing terms only work up to and including the second-order approximation for the static solutions of field equations.  The reason appears to be that the gauge-fixing terms with the covariant derivatives $D_{\mu}$ do not fix the gauge of translational symmetry completely for all orders.  The present gauge-fixing terms in $A^{\mu\nu}$ given by (23) are  satisfactory because they work  for higher orders approximation of solutions.  There are small differences between the field equation (21) and that with the previous gauge-fixing terms.\cite{1}  The differences emerge in the second-order approximation of solutions, and are estimated to be smaller than $10^{-7}$ for the perihelion shift of the Mercury and the bending of light by the sun.  Thus, both solutions are consistent with experiments.\footnote{These solutions have also been verified with symbolic computing (using xCoba in xAct by D. Yllanes and J. M. Martin-Garcia).  The author would like to thank D. W. Yang for his help. }  
 
 We calculated the perihelion shift of the Mercury  to the second 
order with the help of $G^{\mu\nu}(r)$ in 
(16) in terms of  spherical coordinates.\cite{17,1}  
The advance of the perihelion for one revolution of the planet equals  that in Einstein's gravity multiplied by a factor, $
 \left[1 - (E_o^2 - m_p^2)/m_p^2 -  2Gm/3p \right], $ where $p=a(1-\ep^2)$,
 $E_o$ and $m_p$ are respectively the constant (relativistic) energy and mass  of the planet.  The orbit parameter $p$ is related to the eccentricity $\ep$ and the major semiaxis $a$.  The second term, $(E_o^2 - m_p^2)/m_p^2 \approx 10^{-7}$ and the third term, $2Gm/3p \approx 10^{-8}$  are much smaller than 1 in the 
 perihelion shift of the Mercury.   Thus, the difference between  the present  unified model and Einstein's gravity is too small to be detected by observations and experiments.\cite{18}  .

The bending of light can be derived from the propagation of
a light ray in geometrical optics in an inertial frame.  Suppose the light
ray propagates in the
presence of the gravitational field,  its path is determined by the eikonal equation of the light ray, which is `formally' the same as the eikonal equation of the `gluon ray' (18).\cite{1} 
Following the usual procedure,~\cite{17}  the deflection of a light ray differs from the  result in Einstein's gravity by a factor,
$  \left[1 - (9 G^{2}m^{2})/R^{2} \right],$
where $R$ denotes the radius of the Sun.  The additional correction term in the bracket   is  about  $ 10^{-10}$, which is too small to be detected in observations of
 the bending of light by the Sun.\cite{18}  
  
Quantum chromogravity is also consistent with   the
red shift and the time dilatation caused by gravity.  The reason is that light ray propagates in a  space-time with the `effective metric tensor' $G^{\mu\nu}$, so that we have (18) and the `effective metric', $ds^2=I_{\mu\nu}dx^{\mu} dx^{\nu},  I_{\mu\nu}G^{\nu\ld}=\de_{\mu}^{\ld}$.  For example,  the red shift
can be derived from  the  eikonal equation for light ray,\cite{17,1} which is the same as that in Eq. (18) with $G^{\mu\nu}$ given by $G^{\mu\nu}=\eta_{\a\b}J^{\a\mu} J^{\b\nu}$.  We obtain the following result for the red shift of frequency due to the sun,
\be
\om = \om_{o}\sqrt{G^{00}}\approx \om_{o}\left(1+\frac{Gm}{r} +\frac{G^2 m^2}{2r^2}\right),
\ee
where the frequency $\om_{o}$ remains constant during the propagation of the light ray.  This result agrees with experiments.\cite{18,19}

Furthermore, based on the gravitational field equation (21) without the gauge fixing term (i.e., $A^{\mu\nu} = 0$), the gravitational quadrupole radiations  can be 
calculated.  
The result for the power emitted per solid angle  can be shown to be consistent with that  obtained in Einstein's gravity to the second-order approximation in $g\phi^{\mu\nu}$.\cite{20,21} 
 
The emergence of  the effective metric tensor $G^{\mu\nu}$ in the geometric-optics limit is a general property of all wave  equations of quarks and gluons.  It appears as if any classical object moves in a `curved space-time' with the metric tensor $G^{\mu\nu} = \e_{\a\b} J^{\a\mu} J^{\b\nu}$, and behaves in a way consistent with the equivalence principle and the experiments.  However, the true underlying space-time of quantum chromogravity is flat.  Consequently, one has the true conservation of the energy-momentum tensor by Noether's theorem and the usual quantization procedure for the gravitational field\cite{2}  can be carried out for QCG  in inertial frames.  

The unified model of chromogravity suggests that the ``curvature of space-time'' revealed by classical tests of gravity appears to be a manifestation
 of the translational gauge symmetry in the classical limit within flat space-time.  In other words, the unified model suggested a new picture of the physical universe in the presence of gravity, especially at short distances and in the quantum world, which is quite different from Einstein's gravity.  Thus, many conventional results related to short distances and quantum properties in astrophysics and cosmology should be re-examined based on the viewpoint of QCG and quantum mechanics.  For example, the gravitational effects in the microscopic world should be derived from Eqs. (21) and (14) rather than from Eq. (16) (which is an approximate result in the classical limit) or the corresponding `effective metric'  $ds^2=I_{\mu\nu}dx^{\mu} dx^{\nu},  I_{\mu\nu}G^{\nu\ld}=\de_{\mu}^{\ld}$.

We stress that the $T(4)$ gauge symmetry dictates the gravitational coupling of $\phi_{\mu\nu}$ to all other fields, and leads to a very simple gravitational interaction, in comparison with that in the conventional formulations of gravity following Einstein's approach.  For example, the graviton self-coupling in the QCG Lagrangian (7) has no more than four-vertex in Feynman diagrams, just like that of Yang-Mills fields.\cite{2}   In contrast, the graviton self-coupling in Einstein's gravity has N-vertex, where N can be arbitrarily large.  Based on computations of Feynman diagrams\cite{2,22} at one-loop level with dimensional regularization,\cite{23} one sees a close similarity to a renormalizable gauge theory.  The result seems to suggest a conjecture that quantum  chromogravity contains a finite number of counter terms.\cite{5}  The significance of a satisfactory quantum gravity is in a more complete understanding of the whole of field theory rather than in its experimental verifications, which probably cannot be done for a long long time.  However, theoretically, we would like to be sure that  observable results in QCD, electroweak theory, and all calculations at a relative low energy \cite{3,5,24,25,26} will not be upset by the high-order corrections of the gravitational interaction, since all known particles in nature cannot escape from the gravitational interaction. 
 
  When one reflects on the ultimate unification of interactions in nature based on the generalized Yang-Mills framework, it is very unlikely for gravity  to be further unified with other interactions within a simple Lie group with one single coupling constant at a deeper level and still consistent with experiments.  The reason is because there is a qualitative difference between the gravitational constant $g$ with the dimension of length (related to the translational gauge symmetry)  and the dimensionless coupling constants of electroweak and strong interactions. 

In light of previous discussions,  it appears that the known gravitational, strong and electroweak\cite{27} interactions in nature are all dictated by gauge symmetries in the generalized Yang-Mills framework,\footnote{In the literature,\cite{28,7} one can trace the idea of unification of interactions with gauge symmetry to the original discussion by Weyl, and the insight of gravity as a Yang-Mills gauge field was discussed by Utiyama.  Thus, it may be fitting to call the generalized Yang-Mills framework the Yang-Mills-Utiyama-Weyl framework.}  as first advocated by Utiyama,\cite{7} $et \ al$ and particularly stressed by Yang.\cite{28}    Furthermore,  the unified model of chromogravity suggests that the  fundamental roles for our understanding of  the universe are played by  the curvature of the (gauge) connection in fiber bundle\cite{29} rather than the space-time curvature in Riemannian geometry or other geometries.

The work was supported in part by the Jing Shin Research Fund of 
UMassD Foundation.
The author would like to thank Leonardo Hsu for his collaboration at earlier stage, he is a  substantial contributor to the space-time symmetry in inertial and non-inertial frames for the generalized  Yang-Mills  framework.  He would also like to thank D. Fine for useful discussions.

\bibliographystyle{unsrt}

\end{document}